\documentclass[preprint,showpacs,preprintnumbers,amsmath,amssymb]{revtex4}

\usepackage{graphicx}
\usepackage{dcolumn}
\usepackage{bm}
\usepackage{hyperref}

\newcommand{\qed}{\nobreak \ifvmode \relax \else
      \ifdim\lastskip<1.5em \hskip-\lastskip
      \hskip1.5em plus0em minus0.5em \fi \nobreak
      \vrule height0.75em width0.5em depth0.25em\fi}

\begin{document}

\preprint{}

\title{A Further (Itakura-Saito/$\beta=0$) Bi-stochaticization and Associated 
Clustering/Regionalization of the 3,107-County 1995-2000 U. S. Migration Network}
\author{Paul B. Slater}%
\email{slater@kitp.ucsb.edu}
\affiliation{%
University of California, Santa Barbara, CA 93106-4030\\
}%
\date{\today}

\begin{abstract}
We extend to the $\beta$-divergence (Itakura-Saito) case $\beta =0$, the comparative bi-stochaticization analyses--previously conducted (arXiv:1208.3428) for the (Kullback-Leibler) $\beta=1$ and (squared-Euclidean) 
$\beta = 2$  cases--of the 3,107-county 1995-2000 U. S. migration network. 
A heuristic, "greedy" algorithm--using the $\beta=1$ results as an initial configuration--is devised.
While the largest 25,329 entries of the 735,531 non-zero entries of the bi-stochasticized table--in the $\beta=1$ case--are required to complete the widely-applied two-stage (double-standardization and strong-component hierarchical clustering) procedure, 105,363 of the 735,531 are needed (reflective of greater uniformity of entries) in the $\beta=0$ instance. The North Carolina counties of Mecklenburg 
(Charlotte) and Wake (Raleigh) are considerably relatively more cosmopolitan in the 
$\beta=0$ study. The Colorado county of El Paso (Colorado Springs) replaces  the Florida Atlantic county
of Brevard (the "Space Coast") as the most cosmopolitan, with Brevard becoming the second-most. Honolulu County splinters away from the other four (still-grouped) Hawaiian counties, becoming the fifth most cosmopolitan county nation-wide. The five counties of Rhode Island remain intact as a regional entity, but the eight counties of Connecticut fragment, leaving only five counties clustered.
\end{abstract}


\maketitle
We continue our comparative investigations of bi-stochaticizations of weighted, directed networks--in particular, the network of 1995-2000 migration flows between 3,107 U. S. counties--and their associated clustering/regionalization properties \cite{Comparative}.

We have previously "bi-stochasticized" the $3,107 \times 3,107$ matrix of flows by minimizing each of two forms of $\beta$-divergence. "The $\beta$-divergence is a family of cost functions parameterized by a single shape parameter $\beta$ that takes the (squared)-Euclidean distance, the Kullback-Leibler divergence and the Itakura-Saito divergence as special cases ($\beta=2,1,0$ respectively)" \cite{fevotte}.

We--in the extensive series \cite{PBSPNAS,dubes} of applications of the two-stage 
(double-standardization \cite{mosteller}, followed by strong-component hierarchical clustering \cite{tarjan2}) algorithm--had always employed the well-established Kullback-Leibler-based procedure 
($\beta=1$) for double-standardization \cite{knight}. In \cite{Comparative},
we, for the first time, implemented the $\beta=2$ approach \cite{WLK,WLK2}, and found strong differences between the $\beta=2$ and $\beta= 1$ results. In particular, in the $\beta=2$ case, there were 2,707 entries of the associated doubly-stochastic matrix equal to the (maximum possible value of) 1, while in the doubly-stochastic matrix for $\beta =1$, there was only a single such entry.

Here, we seek to expand this pair of analyses to also include the $\beta=0$ (Itakura-Saito) case. Not being aware of any specific effective algorithm for this purpose \cite[p. 357]{WLK2},  we developed a heuristic "greedy" procedure. It relies upon the availability 
(as a starting point) of the previous results of the $\beta = 1$ bi-stochaticization. 

We proceed by randomly choosing a pair ($m_{ij}, m_{kl}$) of the 735,531 non-zero entries in the original data (flow) table. If $i \neq k$ and $j \neq l$, then we ask if $m_{il}$ and $m_{kj}$ are also non-zero. If so (which occurs about $9.22\%$ of the time), we seek that (arbitrarily-signed) value of $x$ which when added to $m_{ij}$ and $m_{kl}$ and subtracted from $m_{il}$ and $m_{kl}$ minimizes the (Burg-entropy-based \cite[Table 2.1]{dhillon}) objective function
\begin{equation} \label{crossratio}
\frac{m_{ij}}{s_{ij}+x}-\log{\frac{m_{ij}}{s_{ij}+x}}+\frac{m_{kl}}{s_{kl}+x}-\log{\frac{m_{kl}}{s_{kl}+x}}+\frac{m_{il}}{s_{il}-x}-\log{\frac{m_{il}}{s_{il}-x}}+\frac{m_{kj}}{s_{kj}-x}-\log{\frac{m_{kj}}{s_{kj}-x}},
\end{equation}
where we impose the constraints, $0 <s_{ij}+x <1, 0 <s_{kl}+x <1, 0 <s_{il}-x <1, 0 <s_{kj}-x <1$.
Here the $s$'s, initially, are chosen to be the corresponding entries of the $\beta=1$ bi-stochasticized table, previously obtained (using the well-known Sinkhorn-Knopp iterative algorithm \cite{knight}). Then, the four indicated entries are updated by either adding or subtracting the optimal value of $x$. This procedure, importantly, preserves the bi-stochasticity of the $\beta=1$ bi-stochastic table from which we have started our heuristic, "greedy" procedure.

The initial value of the sum
\begin{equation}
\Sigma_{i,j}^{n=3107} \Big(\frac{m_{ij}}{s_{ij}} -
\log{\frac{m_{ij}}{s_{ij}}} -1 \Big),
\end{equation}
which is taken (thus, avoiding singularities) only over the 735,531 non-zero entries ($m_{ij}>0$) of the $3,107 \times 3,107$ table, was $4.71219\times 10^{11}$.
Implementing the minimization operation (\ref{crossratio}) 82 million times, and updating the values of the $s$'s as we proceed, we reduced this sum to $1.59538 \times 10^{11}$. 
(On the other hand, the objective function--the Kullback-Leibler divergence [that is, $x \log{\frac{y}{x}} +y -x$]--in the $\beta=1$ case, achieving a minimum value of $4.92974\times 10^8$ there, increases to $5.01181 \times 10^8$, if the 
$\beta=0$ bi-stochastic table, derived from the $\beta=1$ table as its starting point, is substituted in the objective-function calculation.) To indicate the strong convergence of the algorithm, the objective function after 80 million iterations was $1.59541 \times 10^{11}$.

Next, applying the strong-component hierarchical clustering step 
\cite{tarjan2} of the two-stage algorithm \cite{PBSPNAS,dubes}--with 2,517 non-trivial mergings occurring (2,497 for $\beta=1$)--the largest 105,363 entries of the 
$\beta=0$ table were required to complete the clustering, while only 25,239 were needed in the 
$\beta=1$ case \cite{DendrogramRegionalization}. (This appears to be indicative of the greater uniformity of entries in the $\beta=1$ analysis. The $\beta =2$ case, on the other hand,  did not seem to lend itself meaningfully to the application of the hierarchical clustering procedure, due to the large concentration [$87.1284\%$] of its non-zero entries equalling 1, as well as  its relatively small number [57,153 {\it vs.} 735,531] of strictly non-zero entries.)

We, now, present the (ordinally-ranked) dendrogram associated with the $\beta=0$ analysis, while its $\beta=1$ counterpart can be viewed in 
\cite{DendrogramRegionalization}.  The North Carolina counties of Mecklenburg 
(Charlotte) and Wake (Raleigh) are considerably relatively more cosmopolitan in the 
$\beta=0$ study than in the $\beta=1$ analysis, as well as Franklin County, Ohio (Columbus, the state capital). The Colorado county of El Paso (Colorado Springs) replaces  the Florida Atlantic county
of Brevard (the "Space Coast") as the most cosmopolitan, with Brevard becoming the second-most. Only five of the eight counties of Connecticut are clustered in the $\beta=0$ analysis, while all eight
 form a well-defined region in the $\beta=1$ case. The five counties of Rhode Island are grouped in both studies, but the fifth county (Honolulu) of Hawaii is now omitted from the state grouping in the $\beta=0$ study, becoming the fifth most cosmopolitan nation-wide.

Some "fine-tuning" of our clustering results may be subsequently reported, as we continue to run our algorithm, obtaining ever-increasing degrees of the already high convergence already achieved.

We are also exploring the use of additional forms of Bregman divergences--such as the inverse ($\frac{1}{x}$) type \cite[Table 2.1]{dhillon}.

\section{County-Level Dendrogram}
\includegraphics[page=1,scale=.95]{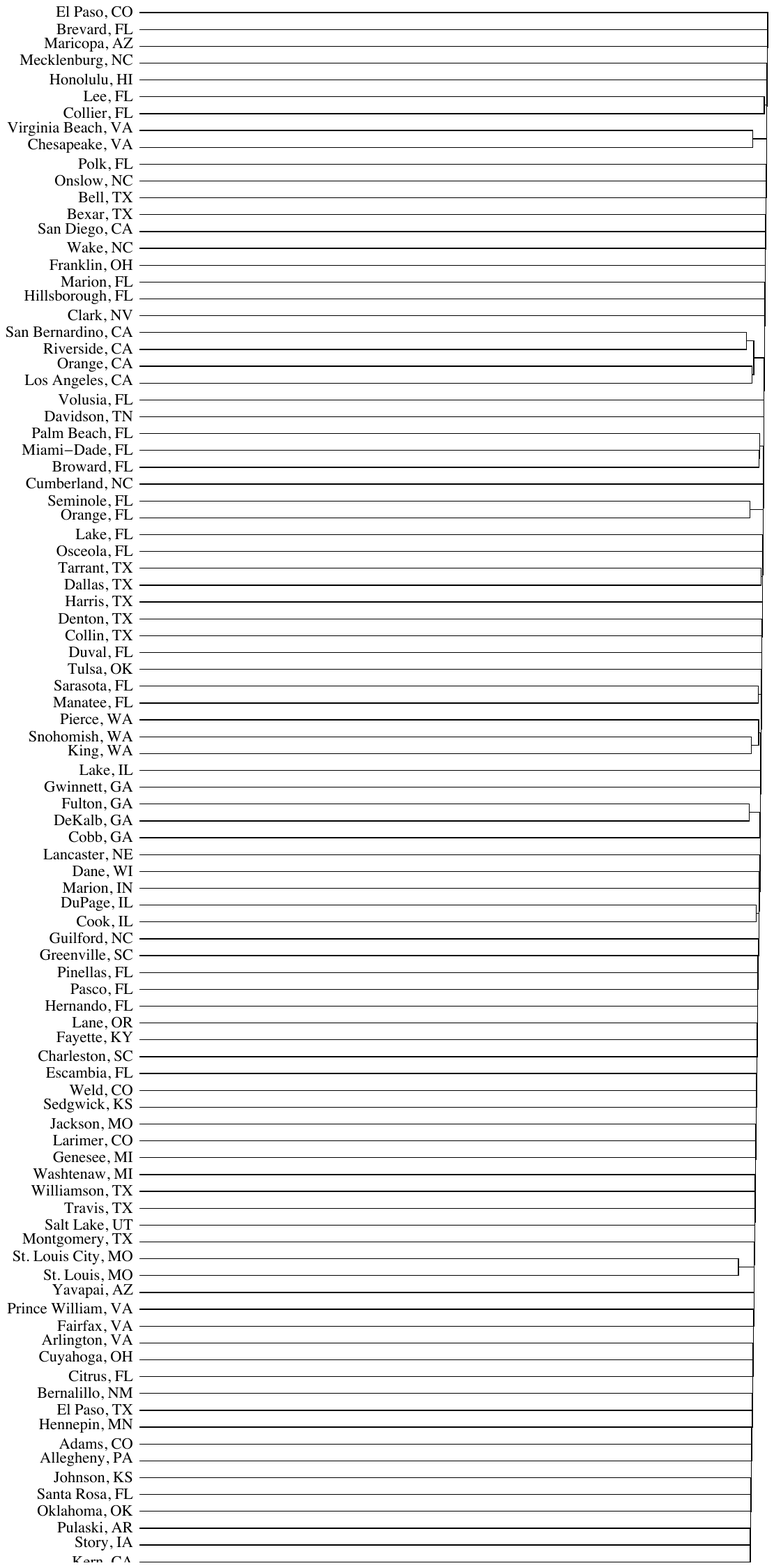}

\includegraphics[page=2,scale=.95]{ItakuraSaitoOrdinalDendrogram2.pdf}

\includegraphics[page=3,scale=.95]{ItakuraSaitoOrdinalDendrogram2.pdf}

\includegraphics[page=4,scale=.95]{ItakuraSaitoOrdinalDendrogram2.pdf}

\includegraphics[page=5,scale=.95]{ItakuraSaitoOrdinalDendrogram2.pdf}

\includegraphics[page=6,scale=.95]{ItakuraSaitoOrdinalDendrogram2.pdf}

\includegraphics[page=7,scale=.95]{ItakuraSaitoOrdinalDendrogram2.pdf}

\includegraphics[page=8,scale=.95]{ItakuraSaitoOrdinalDendrogram2.pdf}

\includegraphics[page=9,scale=.95]{ItakuraSaitoOrdinalDendrogram2.pdf}

\includegraphics[page=10,scale=.95]{ItakuraSaitoOrdinalDendrogram2.pdf}

\includegraphics[page=11,scale=.95]{ItakuraSaitoOrdinalDendrogram2.pdf}

\includegraphics[page=12,scale=.95]{ItakuraSaitoOrdinalDendrogram2.pdf}

\includegraphics[page=13,scale=.95]{ItakuraSaitoOrdinalDendrogram2.pdf}

\includegraphics[page=14,scale=.95]{ItakuraSaitoOrdinalDendrogram2.pdf}

\includegraphics[page=15,scale=.95]{ItakuraSaitoOrdinalDendrogram2.pdf}

\includegraphics[page=16,scale=.95]{ItakuraSaitoOrdinalDendrogram2.pdf}

\includegraphics[page=17,scale=.95]{ItakuraSaitoOrdinalDendrogram2.pdf}

\includegraphics[page=18,scale=.95]{ItakuraSaitoOrdinalDendrogram2.pdf}

\includegraphics[page=19,scale=.95]{ItakuraSaitoOrdinalDendrogram2.pdf}

\includegraphics[page=20,scale=.95]{ItakuraSaitoOrdinalDendrogram2.pdf}

\includegraphics[page=21,scale=.95]{ItakuraSaitoOrdinalDendrogram2.pdf}

\includegraphics[page=22,scale=.95]{ItakuraSaitoOrdinalDendrogram2.pdf}

\includegraphics[page=23,scale=.95]{ItakuraSaitoOrdinalDendrogram2.pdf}

\includegraphics[page=24,scale=.95]{ItakuraSaitoOrdinalDendrogram2.pdf}

\includegraphics[page=25,scale=.95]{ItakuraSaitoOrdinalDendrogram2.pdf}

\includegraphics[page=26,scale=.95]{ItakuraSaitoOrdinalDendrogram2.pdf}

\includegraphics[page=27,scale=.95]{ItakuraSaitoOrdinalDendrogram2.pdf}

\includegraphics[page=28,scale=.95]{ItakuraSaitoOrdinalDendrogram2.pdf}

\includegraphics[page=29,scale=.95]{ItakuraSaitoOrdinalDendrogram2.pdf}

\includegraphics[page=30,scale=.95]{ItakuraSaitoOrdinalDendrogram2.pdf}

\includegraphics[page=31,scale=.95]{ItakuraSaitoOrdinalDendrogram2.pdf}

\includegraphics[page=32,scale=.95]{ItakuraSaitoOrdinalDendrogram2.pdf}

\includegraphics[page=33,scale=.95]{ItakuraSaitoOrdinalDendrogram2.pdf}

\includegraphics[page=34,scale=.95]{ItakuraSaitoOrdinalDendrogram2.pdf}

\begin{acknowledgments}
I would like to express appreciation to the Kavli Institute for Theoretical
Physics (KITP)
for computational support in this research. 
\end{acknowledgments}

\bibliography{ItakuraSaito2}

\end{document}